# Wireless Connectivity in the Sub-THz Spectrum: A Path to 6G

*ABSTRACT*—Wireless communication in millimetre wave bands, namely above 20 GHz and up to 300 GHz, is foreseen as a key enabler technology for the next generation of wireless systems. The huge available bandwidth is contemplated to achieve high data-rate wireless communications, and hence, to fulfil the requirements of future wireless networks. In this paper, we discuss and illustrate new paradigms for the sub-THz physical layer, which either aim at maximizing the spectral efficiency, minimizing the device complexity, or finding good trade-off. The solutions offered by appropriate modulation schemes and multi-antenna systems are assessed based on various potential scenarios.

**From the collaboration between four French institutions**

| **CEA-Leti** | **Centrale-Supélec** | **SIRADEL** | **ANFR** |
|---|---|---|---|
| Simon Bicaïs | Majed Saad | Yoann Corre | Emmanuel Faussurier |
| Jean-Baptiste Doré | Mohammad Alawieh | Grégory Gougeon | |
| | Faouzi Bader | | |
| | Jacques Palicot | | |

**Supported by the ANR-BRAVE project**

# CONTENT

# I. INTRODUCTION

Next-generation wireless networks are imagined to be faster (1 Tbps), more reactive (sub-ms latency), ultra-reliable and denser, to provide for very accurate positioning, highly immersive experiences, smarter autonomous objects, *etc*. Therefore, the exploitation of new and wider bandwidths at higher frequencies is an obvious and promising solution towards 100+ Gbps data rates. With an unprecedentedly large amount of unused bands, the *sub-TeraHertz* (sub-THz) spectrum from 90 to 300 GHz, and the *TeraHertz* (THz) spectrum above 300 GHz, are definitively identified as key enablers for beyond 5G (B5G) communication systems. An aggregated bandwidth of about 105 GHz [1] is possibly available for terrestrial radio-communications between 90 and 275 GHz. Several key applications with strong technological, societal and economical potential impacts are [1]: high speed high-capacity back/front-haul; broadband ultra-low-latency device-to-device (D2D) connectivity; short-range hot-spots; and improved physical layer confidentiality.

SIRADEL (service & software solution provider), CEA-Leti (research institute), Centrale-Supélec (engineering school), and ANFR (spectrum regulator) have launched a partnership in 2018 to elaborate and study new signal waveforms operating in the future 6th-generation (6G) sub-THz spectrum. This white paper gives an overview of the studies performed in this framework.

## *Challenges*

Promises are great, yet many challenges need to be addressed to achieve high data-rate communication in sub-THz frequencies. First of all, due to the strong obstruction losses, the connectivity link is mainly restricted to line-of-sight (LoS) situations, slightly obstructed line-of-sight (OLoS), or requires a strong reflection. The free-space propagation losses increase with the square of the frequency as depicted in Fig. 1, but it could be compensated by using high gain antennas. Severe constraints on antenna directivity and alignment are entailed.

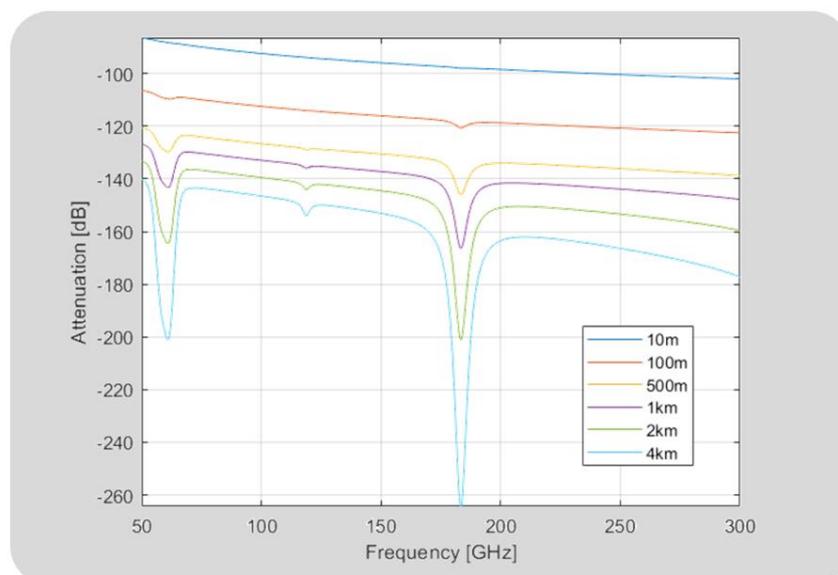

Fig. 1: RF signal attenuation as a function of the carrier frequency assuming an atmospheric water vapor density of 7.5g/m$^3$.

In addition, new physical (PHY) layer algorithms are required on the digital side to achieve a high data rate.

Traditional techniques cannot be directly transposed to sub-THz bands as they do not consider the specific radio-frequency (RF) impairments of sub-THz systems. In particular, sub-THz systems suffer from medium to strong oscillator phase noise (PN)[1]. Therefore, the study of new digital transmission schemes optimized to mitigate those RF impairments is necessary to guarantee good performance [2].

Besides, multiple-input multiple-output (MIMO) systems in sub-THz bands are not mature enough compared to existing techniques deployed for cellular networks. In general, it is important to reduce the large power consumption experienced by power amplifiers (PA) at each RF chain due to high peak-to-average power ratio (PAPR), and hence, to develop techniques that achieve a high Spectral Efficiency (SE) with a given power expenditure – see for example [3] [4] where it is highlighted that the power-efficient modulation with Index Modulation consumes less power to achieve the same SE.

The available transmit power at sub-THz bands is limited (generally in order of 20-30 dBm), which makes it even more crucial to operate the power amplifier at a higher efficiency. Thus, the low-PAPR single carrier (SC) modulations are more suitable, especially as they are compliant with sub-THz channel characteristics [5].

Hence, the wireless Tbps requires revisiting the waveform design and the naïve adopted techniques (e.g., using higher order modulation) to achieve green beyond-5G communications [3].

Finally, semiconductor technologies must be enhanced to meet the expected performance in terms of SE [6] but at a reasonable cost.

## *Proposed solutions*

First, this white paper describes three different scenarios in which the use of high-capacity frequency bands could give new major opportunities, namely, outdoor backhaul, short-range, and D2D communications. The sub-THz design limitations are cross-correlated to those contemplated applications to determine the characteristics of adequate future physical layers.

Second, for each envisaged scenario, namely, *high capacity backhaul, enhanced short-range hotspot*, and *device-to-device communications,* the physical layer is presented, and the system performance is assessed considering realistic environments (digital twins) as well as RF impairments models. Eventually, a synthesis of the key performance indicators (KPI) for the different scenarios is presented.

The proposed solutions have been developed in the frame of BRAVE project funded by the French research agency ANR. This project has started in January 2018 and will finish in November 2021. As illustrated in Fig. 2, it is positioned at the earliest stage of the beyond-5G or 6G investigations, with aim to support and possibly influence the emergence of new sub-THz technologies, future standardization work and future regulation studies, by providing adequate PHY-layer models, efficient waveforms, test simulation framework, and scenarios assessment.

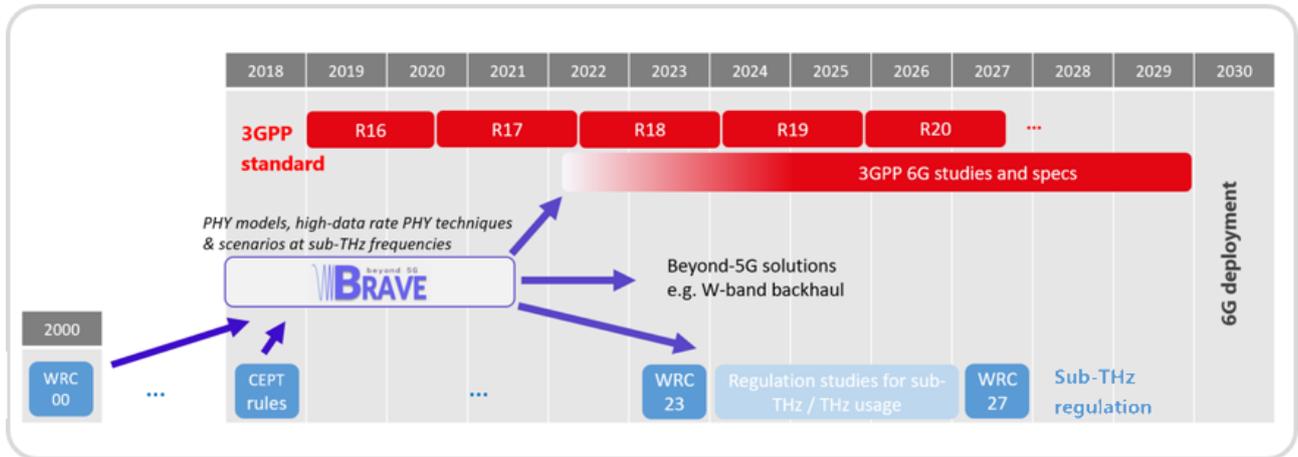

Fig. 2: Positioning of the ANR BRAVE project.

# II. DESIGN LIMITATIONS IN SUB-THZ BANDS

*The physical layer for sub-THz systems*

Sub-THz communication systems are considered as a foremost solution to meet the requirements of beyond 5G (B5G) and 6G networks. Some of the contemplated use cases and applications for sub-THz systems are the following:

a) **High-capacity backhaul**

The envisioned ultra-dense network topology in urban areas or local private networks (e.g., for industry, transport hubs, stadia, or smart cities) with the extreme data-rate, capacity, and latency requirements makes the fiber-based backhauling highly desirable, but sometimes complicated due to current fiber networks penetration (variable from one country to the other) and local installation constraints. The wireless backhaul infrastructure is needed as an alternative or complement to the optical fiber deployment; it offers more agility, shorter installation times, and (in case of a mesh architecture) strong reliability. It may also provide connectivity to mobile or even flying access points. High data-rate wireless backhauling is a valuable competitive technology, which benefits from lower deployment costs and constraints.

b) **Enhanced short-range hotspot**

An ultra-high data rate downlink serving single/multi-user(s), with end-users complexity and energy constraints. Envisaged applications are short-range hotspots delivering high-speed data to demanding applications such as enhanced Wireless LAN (WLAN), kiosk, augmented or virtual reality, connectivity for robots, drones or autonomous fleets that require sub-ms reactivity.

c) **Device-to-device communications**

A symmetric high data-rate link with energy and architecture complexity constraints for D2D communications. It includes inter- or intra-chip communications, wireless connectors, or connection between devices in a server farm.

Other use-cases could be envisaged, such as improved physical layer confidentiality thanks to radio confinement and high antenna directivity. Accurate positioning and high-resolution sensing are also strong motivations for the use of the sub-THz spectrum.

Despite the evolution of semiconductor technologies, more research is required to design new physical layer algorithms and transceiver RF architectures to mitigate the severe RF impairments of sub-THz systems.

*The two sub-THz systems paradigms*

Fig. 3 illustrates the path from the specific sub-THz properties to the proposed paradigm duality: implementation of either a spectral efficient or a low complexity physical layer. Trade-off between spectral efficiency and complexity may also be envisaged for some applications. The recent measurement campaigns have shown that sub-THz propagation channels are largely dominated by a single path (often LoS direct path), which provides most of the energy contribution. It is due to the stronger channel sparsity at those frequencies, in particular in open or urban environments, and to the usage of highly directive antennas, sometimes at both transceiver sides. It follows that single-carrier (SC) communication systems are envisaged to provide low-complexity RF architecture and operate at high-PA efficiency. In particular, sub-THz systems suffer from medium to strong PN impairments, resulting from the poor performance of high-frequency oscillators. The phase impairment severely deteriorates the performance of sub-THz communication systems.

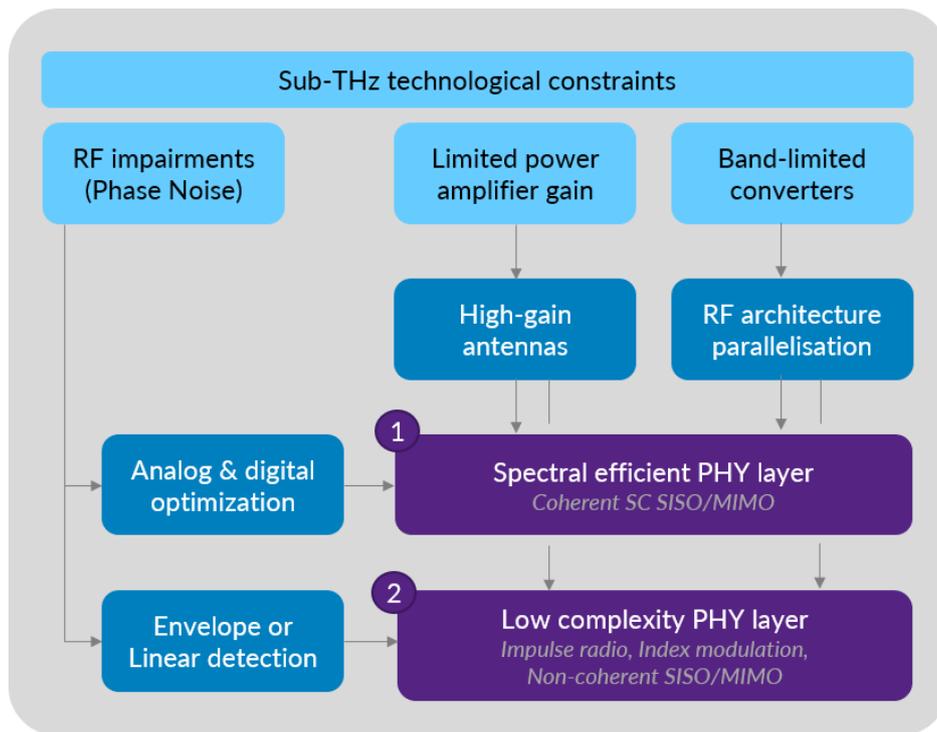

Fig. 3: Illustration of the physical layer paradigms for sub-THz communication systems. The constraints coming from the propagation, namely the need for high-gain antenna and beam alignment processing, are common for the two paradigms.

On the one hand, to mitigate the impact of this major RF impairment, the optimization of the signal processing related to physical layer algorithms is mandatory. For example, the optimized modulation and demodulation schemes can be used to achieve some PN robustness. On the other hand, efforts should be done for the design of new oscillator generation techniques compatible with CMOS technology. Last, delivering high power (and relatively high efficiency) with CMOS compatible technologies is still an open research topic. Alternatively, the use of receivers based on envelope or energy detection is also considered, which enables a frequency down-conversion from passband to baseband without the impact of phase impairments.

In addition, the digitalization of large bandwidth signals entails severe constraints on digital-to-analog converters (DAC) and analog-to-digital converters (ADC). The parallelization of the transceiver RF architecture appears to be essential in order to relax the ultra-high sampling rate constraint on converters that are also more power-hungry and costly. Some of the investigated approaches rely on channel bonding systems, *i.e.,* the analog aggregation of multiple carriers.

With regard to the contemplated applications and the presented constraints related to sub-THz communication systems, two paradigms arise for the physical layer: high data rate versus low complexity. We describe in the following paragraphs the features of these two paradigms.

a) ***Spectral efficient physical layer***

Its physical layer corresponds to a communication system whose objective is to maximize the spectral efficiency such as for high capacity back-haul. This physical layer implies the use of in-phase / quadrature (IQ) transceivers, high-quality RF components, and high-order modulation schemes. Concerning the research on the physical layer for such communication systems, most of the current works investigate the optimization of channel bonding systems and the related signal processing.

b) ***Low-complexity physical layer.***

Conversely, in this paradigm, communication systems aim to minimize the complexity of the architecture or the energy consumption to achieve a given rate. Contemplated applications, in this case, are either the enhanced hot-spot or short-range communications. This paradigm entails a complexity/power-limited regime, and hence using a simple RF architecture, analog or basic (e.g., on-off keying) modulation schemes. Numerous research approaches are under investigation for the development of the physical layer for complexity/energy-constrained systems. Research works include the use of index modulation [7] [8], the design of high data-rate impulse radio, the joint optimization of analog and digital signal processing. It is worth mentioning that Index modulation (IM) provides a relevant solution for power-constrained sub-THz systems, and it also allows to achieve a good trade-off between both paradigms (spectral-efficient and complexity/energy-constrained systems), see [3] [4] and references therein.

> ***We, therefore, claim that a constraint driven design of sub-THz communication systems is essential to the sub-THz paradigm duality: implementation of either a spectral efficient or a very low complexity physical layer.***

# III. SPECTRUM, OPPORTUNITIES AND REGULATION

The sub-THz spectrum between 90 GHz and 300 GHz offers opportunities for huge bandwidths, up to several tens of GHz, which is required to increase data rates and network capacities beyond 5G performance. The aggregation of all possibly allocated sub-THz bands should enable the achievement of the 1-Tbps wireless communication.

The radio spectrum above 90 GHz is today essentially known for being used by scientific services (e.g., astronomy observation, earth exploration satellite services, and meteorology, etc.) and has never been used effectively for radio wireless communications purposes. It should however be noted that the radio regulations already allocated several frequency bands above 90 GHz to the fixed and mobile services.

> *The international Radio Regulation (RR) has decided in 2000 which frequencies between 90 and 275 GHz can be used for fixed or mobile radio-communication services: 105.1 GHz in total. Eighteen years later, the European regulation has published recommendations for fixed sub-THz links, considering the compatibility with the passive services.*

## *Global Radio Regulation*

The Radio Regulations (RR) provide an international framework for the management of radio frequencies up to 3 THz, with agreed frequency allocations stopping at 275 GHz, as shown in Fig. 3.

**Frequency allocations between 90 GHz and 275 GHz were decided at WRC-2000** (World Radiocommunication Conference), within the frame of agenda items 1.16 and 1.17. They provided visibility to stakeholders on available frequencies for a large set of radiocommunication services. ITU-R studies performed at that time indicated that sharing between the EESS (passive) and the fixed service is generally possible in frequency bands with high atmospheric absorption but may require constraints to be applied on the fixed service to protect passive space sensors. With respect to the protection of the terrestrial radioastronomy service, it was indicated that national administrations may have to establish coordination zones around millimetre-wave astronomical observatories. Coordination radii in the order of 100 km may be necessary.

Frequencies that may be used by active radio systems are bordered by passive bands, i.e., bands covered by footnote RR 5.340 which prohibit any emission.

Around 15 GHz of spectrum is available in the W-band, and more than 30 GHz of spectrum is available in the D-band. The total allocated bandwidth in the sub-THz spectrum is 105.1 GHz.

White paper - Wireless connectivity in the sub-THz spectrum: A path to 6G

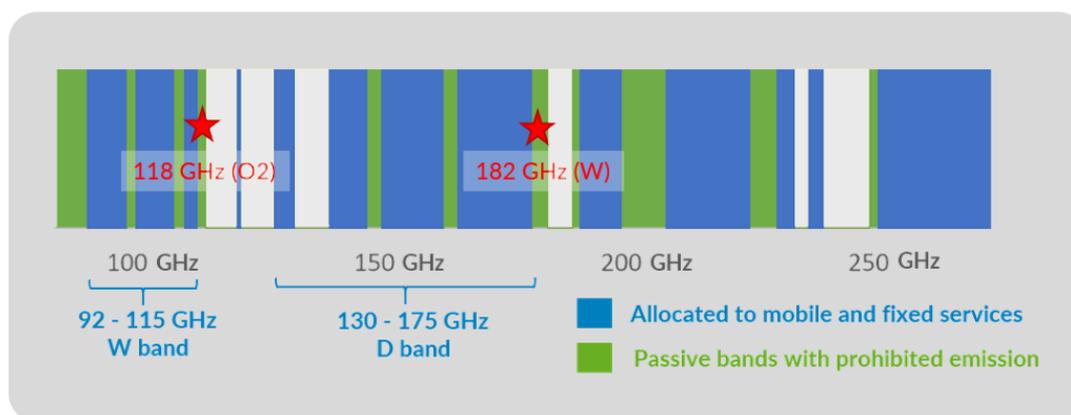

Fig. 4: Frequency allocation to active communication or passive services by international regulation in the spectrum from 90 to 275 GHz.

Effective regulatory solutions at European level should be based on RR frequency allocations. Their development and national implementation are driven by market demand. **European harmonisation measures** related to the fixed service have been completed in 2018, while radiodetermination applications in the range 120-260 GHz are considered in ongoing activities.

*CEPT regulations for the fixed service*

CEPT (European Telecommunication Office) investigations on FS (Fixed Services) in the frequency ranges 92–114.25 GHz (W-band) and 130–174.8 GHz (D-band) were completed in 2018 with the publication of the four deliverables described in Table 1.

| CEPT publication | |
| --- | --- |
| **ECC Report 282 [9]** | Point-to-Point Radio Links in the Frequency ranges 92-114.25 GHz and 130-174.8 GHz |
| **Revised ECC Recommendation (14)01 [10]** | Radio frequency channel arrangements for fixed service systems operating in the band 92-95 GHz |
| **ECC Recommendation (18)01 [11]** *W-band* | Radio frequency channel/block arrangements for Fixed Service systems operating in the bands 92-94 GHz, 94.1-100 GHz, 102-109.5 GHz and 111.8-114.25 GHz |
| **ECC Recommendation (18)02 [12]** *D-band* | Radio frequency channel/block arrangements for Fixed Service systems operating in the bands 130-134 GHz, 141-148.5 GHz, 151.5-164 GHz and 167-174.8 GHz |

Table 1: List of CEPT publications related to European harmonization rules in sub-THz spectrum.

**The main CEPT remarks on sub-THz fixed services in W- and D-bands** can be summarized as:
1. Very wide bandwidths allow for low-cost multi-service operations;
2. Deployments will be easier compared to wired alternative solutions;
3. High security is ensured by lower risks on interference or signals capture.

**If we look at the details…** The channel arrangements in ECC/REC/(18)01 and ECC/REC/(18)02 account for flexible FDD/TDD deployment, with continuous frequency bands of 250 MHz, without specifically defining either paired or unpaired use. There is no strict limitation on the number of aggregated bands, i.e., no limitation on the maximum channel bandwidth. Bands and Carrier Aggregation (BCA) may also be considered for the W-band and D-band to improve capacity and link availability.

ECC/REC/(18)02 states that optimised trade-off between very wide channels and spectrum efficiency would allow achieving very compact equipment and low power consumption for ultra-high-capacity backhauling, front-hauling, and possibly fixed wireless access, with up to about 1 km hop lengths in line-of-sight conditions. High-density short links under 200 m can be used for 5G mobile backhauling tail link with capacity of over 10 Gbit/s. Indoor application for internal connection of a data centre (inter-server) is also considered with short links in the order of tens of metres, providing capacity around 40 Gbit/s. The report also mentions link capacity in the order of 100 Gbit/s.

For D-band systems simulation, ECC Report 282 considers maximum antenna gain of up to 40 dBi and +5 dBm output power.

The conclusions and methodologies in ECC Report 124 have been extrapolated in ECC/REC/(18)01 (D-band) and ECC/REC/(18)02 (W-band) in order to establish the FS unwanted emission mask to ensure the protection of EESS (passive). This approach had already been applied to the FS band 92-95 GHz in ECC/REC/(14)01. The unwanted emissions of FS transmitters that are falling into adjacent passive bands should be limited at the FS antenna port, as shown in Fig. 5.

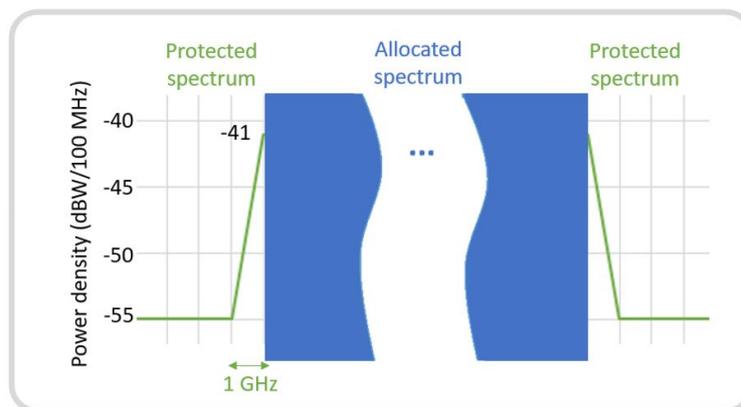

*Fig. 5: Unwanted emission mask for fixed services as recommended by CEPT.*

Finally, it is recommended the calculation of appropriate separation distances between FS transmitters and a radio-astronomy site are done on a case-by-case basis by the national administration.

### *Radiodetermination applications in 120–260 GHz*

Work has been initiated in 2019 within CEPT to address the spectrum requirements on radiodetermination application (determination of position, velocity, etc) within the frequency range 120 GHz to 260 GHz, leading to ETSI report [13]. This document identifies the need for additional frequencies allocated to Ultra-Wideband (UWB) technology and to different measurement applications which cannot be conducted adequately at the moment in the SRD (Short Range Devices) bands 122-123 GHz and 244-246 GHz due to the limited bandwidth. It distinguishes between applications of type A, B or C. Those of type A may radiate maximum power in any direction into the open sky, which would be most critical regarding the generation of harmful interference into radiocommunication services. Other applications are less critical in terms of interference as they are emitting towards the ground (type B) or inside a confined environment (type C).

In terms of regulatory approach, the reference to UWB basically suggests that the possible spectrum regulations to be developed would not be limited within the relevant spectrum allocations of the Radio Regulations and that no protection can be granted to these radiodetermination applications. Administrations would have to assess the justification for these requirements, to consider some enforcement aspects in relation with the various specific applications, and also to consider suitable mechanisms to ensure the protection of radiocommunication services e.g., by ensuring that a required geographical separation is met in case of in-vehicle sensor applications.

# IV. SYSTEM MODEL

In the remaining of this document, the aforementioned dual paradigm is illustrated based on three applications (backhaul, short-range indoor communication, and D2D) having different requirements. Appropriate physical layer solutions are suggested and evaluated from a realistic sub-THz system model.

The system models developed for evaluation of traditional sub-6GHz physical layers have to be reconsidered with regard to specific sub-THz characteristics and scenarios. In particular, the propagation channel properties, the antenna systems, and impairments at Tx/Rx RF chains do critically affect the link performance and the service coverage range. The sub-THz propagation differs from the one at lower frequencies by severe attenuations due to walls, windows, foliage, vehicles, or bodies, and very limited diffraction. The main propagation mechanisms are the LoS direct path and the reflections free of any obstruction. When associated with directive antennas, the propagation channel often becomes restricted to a single dominant path, leading to a frequency-flat response. Even in the presence of a clear propagation path, the communication range is affected by a path-loss proportional to log(frequency), high atmospheric absorption, and rainfall attenuation.

A fine characterization of the propagation channel is required for a realistic evaluation of new sub-THz systems. Only a few channel-sounding campaigns, complex and costly, have been published yet [14]. The results of these measurement campaigns can be complemented with numerical predictions that can produce on-demand channel models. In this work, we consider a model based on a deterministic ray-tracing tool for sub-THz frequencies [15]. This tool benefits from a LiDAR point cloud or detailed 3D representations to get a realistic prediction of the blockages and losses due to trees and street or indoor furniture.

Regarding the electronic systems themselves, a critical impact from RF impairments – such as non-linearity, IQ imbalance, and oscillator PN – are expected. The non-linearities of RF front-end analog blocks increase challenges in both modelling of circuits and designing the mitigation techniques. Hence, it is clear that low order Amplitude-Phase Modulations (APMs) and OOK (On Off Keying) based systems are more convenient. We focus in this paper on the impact of PN on system performance and model the PN with an uncorrelated Gaussian process according to [16]. In the following, the three studied scenarios consider sub-THz links in the D-band at 150 GHz with a channel bandwidth of 1 GHz.

# V. HIGH-CAPACITY BACKHAUL

The optical fiber remains the best solution to deliver reliable and fast back/front-haul capacity for new-generation access networks. However, a dense mesh back-haul topology based on a wideband wireless system may be a valuable alternative that allows for lower short-term investment, faster deployment, and agility. Such solutions are already trialled at 60 GHz to feed outdoor small-cell networks with more than 1 Gbps per link. The possibility for transporting much higher capacity in the sub-THz spectrum is obviously a fantastic perspective. As discussed, strong obstructions drastically impact propagation, thus imply deployment in a clear space. We assume here a D-band (at 150 GHz) coherent transceiver deployed in an outdoor environment. Network simulations for analysis and optimization of the cover range and deployment strategies rely on two key components. First, the link performance is extracted from a PHY-layer model such as throughput versus signal-to-noise ratio (SNR) for various levels of RF impairments. Second, the reception quality is assessed from the propagation channel prediction according to various scenarios, and link budgets.

As an application example, we assume an aggregation of multiple SC modulations with 1 GHz bandwidth (800 MHz useful) and a LDPC code. The choice of a channelization of 1 GHz is motivated by current state-of-the-art of analog-to-digital converters. The system performance, using either a classical quadrature amplitude modulation (QAM) or optimized modulation and demodulation schemes [2], is assessed under strong PN condition – with a noise floor spectral density of −100 dBc/Hz. Due to the strong PN level, the achievable SE when using QAM is limited to 2.5 bit/s/Hz for any SNR. In contrast, when optimized schemes are used, the reachable SE is 5.5 bit/s/Hz with a SNR of 30 dB and a strong phase noise (for reference, the maximum reachable SE is 7.2 bit/s/Hz without phase noise). Interested readers may refer to [2] for more details.

Further numerical evaluations consider a city centre environment with medium-density vegetation, where deterministic propagation is computed, while the system performance is characterized for an outdoor mesh back-haul scenario. The sub-THz devices are supposed to be installed on lampposts, located along the streets, in order to transport the data streams from a fiber point of presence to dense local high-speed AP's (Access Points) that either provide mobile or fixed wireless access typically at a lower frequency. The antenna beam at each device is supposed to be perfectly aligned on the dominant propagation path, which is either the direct path (LoS or OLoS when obstructed by some foliage), a reflected path, or (rarely) a diffracted path. We assume a transmit power of 30 dBm, a maximum beam antenna gain of 25 dBi, and a noise figure of 12 dB. The SNR is estimated from the propagation paths and mapped to the SE. The resulting throughput as computed around a lamppost-mounted device is illustrated by the heat maps in Fig. 6. They show how the lamppost-to-lamppost link performance depends on the PN level and the selected modulation scheme.

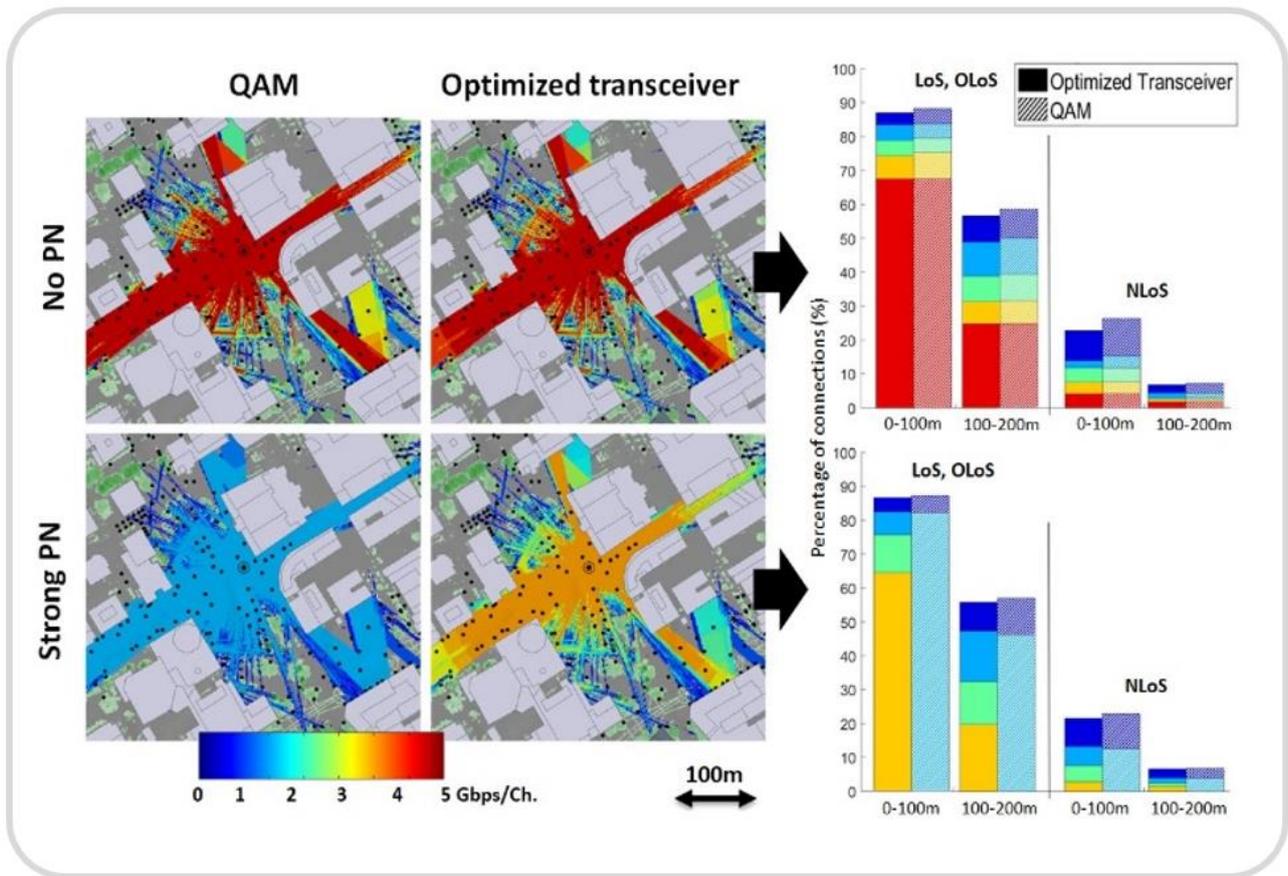

*Fig. 6: Throughput from standard or optimized modulation scheme, with or without phase noise: Heatmap around a back-haul device (left) and Statistics from several back-haul links (right).*

The connectivity is strongly affected by building and foliage obstructions, while LoS propagation and strong clear reflections lead to the best performance. Propagation is computed for more than 1800 lamppost-to-lamppost links in the city centre [17], leading to the connectivity statistics presented in Fig. 6 for LoS/OLoS situations and two different ranges, either [0;100[ meters or [100;200[ meters. These throughput heat maps and statistics demonstrate how significant the gain from sub-THz optimized modulation and demodulation schemes is. The NLoS connectivity statistics also indicate that in some circumstances, a sub-THz link may occur beyond building obstruction; this could offer additional path diversity in a mesh network, e.g., for routing optimization or better protection.

A network infrastructure has been designed based on those assumptions and realistic propagation constraints in order to connect FWA (Fixed Wireless Access) customer equipment in a North-American residential scenario with high vegetation density. Actually, both the multi-hop backhaul and access links are supposed to use D-band spectrum to provide a peak-hour 100 Mbit/s downlink average data rate to half the houses in the environment. A less efficient system is considered for the access link, as the access antenna is covering a whole sector, and the equipment at the customer side (installed on the house's façade) must be of reasonable cost; the resulting link power budget is 27 dB less favorable. One possible infrastructure (obtained from radio-planning techniques) is shown in Fig. 7 (left). The network is composed of 4 fiber points of presence (PoP) and 57 sub-THz wireless nodes (35 nodes with both access and backhaul functions; and 22 additional backhaul nodes) for an area of 0.5 km². The most loaded access and backhaul links do respectively use 5 and 8 channels of 1 GHz bandwidth to deliver the considered data traffic. More details and deployment options are described in [18].

A similar study is applied to an indoor mesh backhaul network [17]. The deployment of 7 mesh nodes in one floor of a 9600 m² wide shopping mall, as depicted in Fig. 6 (right), is shown to offer broadband connectivity to several distributed fixed relays.

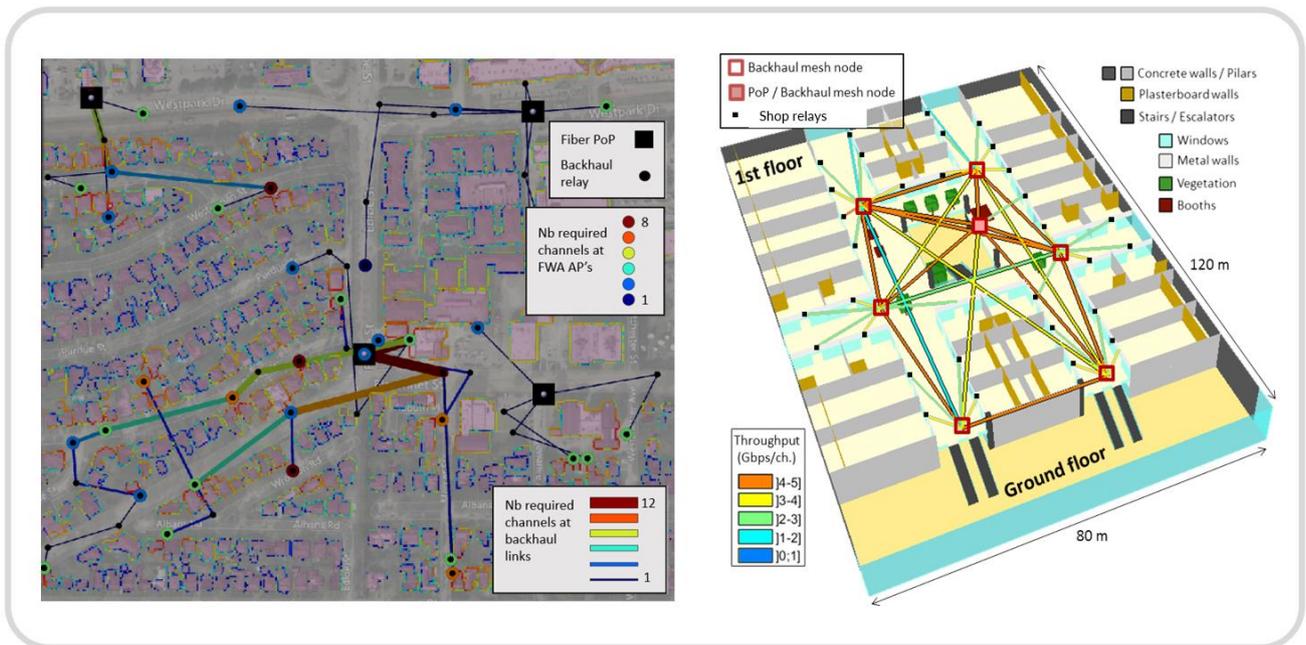

*Fig. 7: Demonstration of dense wireless infrastructures based on D-band backhauling and optimized modulation, for a residential Fixed Wireless Access (left) or a flexible broadband connectivity inside a large venue (right).*

## VI. ENHANCED SHORT-RANGE HOT-SPOT

The wireless ultra-high data rates are also required for many emerging short-range indoor scenarios such as a kiosk, indoor hot spot (high throughput WLAN), server farm, etc..., where the distance between transceivers is limited to few meters. Recently, the IEEE 802.11ax standard in the sub-GHz band enhanced the data rates through the naïve approach of using: higher-order modulation up to 1024 QAM, wider channels, and the adoption of MIMO technologies. However, the sub-THz systems suffer from more severe RF challenges and technological limitations such as low transmit power, PN, non-linearity of PAs (Power Amplifiers), and limited resolution for low power/cost ADCs. Considering these limitations and RF impairments, the usage of high-order APMs becomes not a suitable solution in the current sub-THz technology. In addition, the indoor sub-THz channel characteristics with the short distance and few survival paths allow replacing the multi-carrier waveform with high PAPR in IEEE 802.11ax by power-efficient SC schemes. Based on these sub-THz challenges, we proposed to use a power-efficient low-order SC modulation and then increase its spectral efficiency by Index Modulation (IM) and MIMO techniques [7] [4].

Index Modulation techniques have attracted tremendous attention due to mainly its energy efficiency (EE) and/or SE gain. For instance, generalized spatial modulation (GSM) [19] in the spatial IM domain allows transmitting different symbols simultaneously like MIMO systems with Spatial Multiplexing (SMX), but GSM activates certain Transmit Antenna (TA) set from all the available TAs to convey additional information bits by the index of the activated combination. This indexation technique enhances the SE and EE, and thus it reduces the required modulation order to achieve high throughput systems and mitigate the aforementioned impairments. However, GSM system is sensitive to channel correlation [7] that highly affects its performance, but its optimization/adaption reduces this degradation [20]. Similarly, exploiting the polarization IM dimension as in Dual Polarized GSM (DP-GSM) limits the spatial correlation effect while enhancing more the SE [21].

The study of these candidate systems (GSM, DP-GSM, SMX) with non-coherent optimal detection in sub-THz channels and subjected to RF impairment revealed that (DP)-GSM-QPSK has better performance, lower power consumption and complexity, higher robustness to PN, and few-bits ADC resolution requirement compared to SMX-QAM. Note that the (DP)-GSM systems mainly have three drawbacks: (i) larger TA array where only a subset transmits APM symbols (inefficient exploitation of the available spatial resources); (ii) higher transmitter cost with full-RF architecture, which is needed to avoid the SE degradation due to pulse shapes time-truncation at symbol rate; (iii) the PAPR increases in GSM with full-RF transmitter architecture due to transmitting zero symbols on non-activated antennas [3]. However, large TA array is feasible at these frequencies due to the small antenna size. Similarly, GSM transmitter cost is affordable and realistic in the downlink scenario, where the access-point and kiosk machine act as transmitters. Besides, the reduced power amplifier efficiency due to the PAPR increase is counterbalanced by the significant performance gain of (DP)-GSM-QPSK that makes GSM a potential sub-THz candidate [3].

Consequently, these results and GSM drawbacks motivate us to propose a novel IM domain [8], "Filter IM domain" that provides another novel degree of freedom and allows reaching even higher SE/EE in SISO prior to its MIMO exploitation. This filter IM domain generalizes most existing conventional systems (OOK, Pulse Position Modulation (PPM), linear modulations, Nyquist and faster-than-Nyquist systems, etc.) and SISO IM domains (e.g., frequency, time IM domains). Within the filter IM domain, a novel modulation scheme called Filter Shape Index Modulation (FSIM) is proposed in SISO [8] and MIMO [4], to exploit all available time/frequency and spatial resources in contrast to time/frequency and spatial IM domains. The pulse-shaping filter is indexed and changed at each symbol period to convey additional information bits to those transmitted in the APM data symbols. The FSIM system, even with non-optimal filters, achieves significant SE/EE gains in AWGN, frequency selective, and flat fading channels [8].

Striving for further SE and EE improvement, the proposed filter IM domain is extended to MIMO as depicted in Fig. 8 to exploit IM and multiplexing gains, and thus reach the highest SE gain (Fig. 7-8 in [4]). The proposed SMX-FSIM system conveys information bits by multiplexing different APM symbols simultaneously and performing independent parallel filter shape indexation at each TA. Moreover, the proposed SMX-FSIM system is explored with a linear receiver architecture and independent parallel detectors to achieve lower complexity and faster processing.

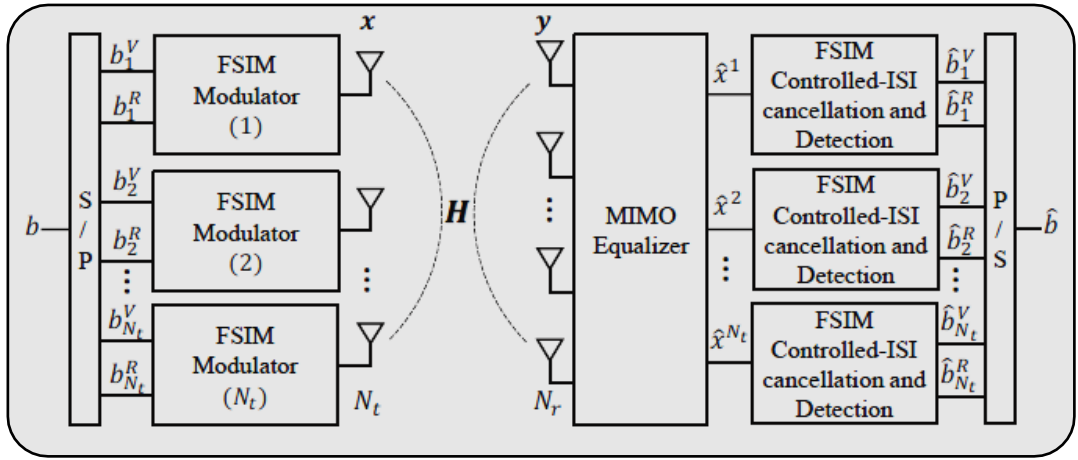

Fig. 8: System model for MIMO SMX-FSIM.

The evaluation results reveal that the low-power wireless Tbps system can be enabled even with current electronics sub-THz limitations by using power-efficient low-order SC modulation accompanied with MIMO and spectral-efficient IM (e.g., MIMO (DP)-GSM-QPSK [7] [21], MIMO SMX-FSIM [4]). However, the latter SMX-FSIM does not provide only better SE/EE but also overcomes the previously highlighted drawbacks of the other most competitive candidates (DP)-GSM accompanied with several advantages in: performance, robustness to phase noise, transceiver cost, SE/EE gain, power consumption, and system flexibility at the cost of a limited computational complexity linear increase with the filter length as shown in Table 2 [4]. This high flexibility allows using a reconfigurable based access-point architecture based on SMX filter IM domain to serve different devices with variable constraints (e.g., SMX-FSIM for high SE/EE, and its special case SMX-OOK for extreme low complexity devices).

White paper - Wireless connectivity in the sub-THz spectrum: A path to 6G

Therefore, all the analysis leads to promote MIMO SMX-FSIM as a very competitive and promising candidate for low-power wireless ultra-high data rates system in sub-THz bands.

Table 2: Summary of different MIMO systems targeting low-power wireless ultra-high data rates.

|  | SMX-QAM | GSM | SMX-FSIM **Most promising solution** |
|---|---|---|---|
| **Spectral efficiency** (Fig. 7-8 in [4]) | Medium $N_t \log_2 M$ | Medium $N_a \log_2 M + \log_2 \binom{N_t}{N_a}$ | High $N_t (\log_2 M + \log_2 N)$ |
| **Robustness to PN** (Fig. 10-11 in [4]) | Low to Medium | Low to High (High with ML) | High |
| **PAPR** (Fig. 12 in [4]) | Low | Low to High (Depends on $M, N_a, N_t$) | Medium |
| **Energy efficiency** (Table 5 in [4]) | Medium | Medium | High |
| **Linear detector complexity** (Table 4 in [4]) | Medium | Low | Medium to High (Depends on filter length) |
| **Cost based on nb of RF chains** (Section IV-C in in [4]) | Low to Medium | High | Low |
| **Flexibility (reconfigurable system)** | Low | High | High |

*A system based on spectral-efficient Index Modulation with power-efficient low-order modulation is very promising for low-power wireless ultra-high data rates, because it provides higher SE/EE and relaxes the most critical requirements of sub-THz/THz hardware (e.g., PN, linearity, transmit power/SNR, etc.).*

# VII. DEVICE-TO-DEVICE COMMUNICATIONS

For this last scenario, we assume a D2D communication with complexity and energy constrained transceivers. To achieve high throughput communications with low-complexity low-power transceivers, we consider MIMO systems with energy detection (ED) receivers as depicted in Fig. 9. Combining spatial multiplexing with ED receivers enables to achieve high throughput and PN robust communications with a simple RF architecture. In details, each spatial stream is transmitted with a directive antenna on a LoS channel. Subsequently, there are two major challenges to achieve spatial multiplexing with non-coherent transceivers in sub-THz frequencies. First, the channels are strongly and spatially correlated and the resulting interference is nonlinear due to the ED at the receiver. In addition, designing a compact system with a low-aperture antenna to limit the interference is particularly challenging for sub-THz frequencies. Nevertheless, the recent work in [22] has demonstrated that spatial multiplexing with non-coherent sub-THz transceivers can be realized on strongly correlated LoS channels while using state-of-the-art antennas and a joint demodulation algorithm.

To demonstrate the benefits of this approach, system simulations were performed. We assume a carrier frequency of 150 GHz, a bandwidth of 1 GHz, an antenna gain of 32 dBi with a beam width of 3°, and sidelobe level at −20 dBi and a transmission range of 5m. The considered MIMO system uses uniform linear arrays of $N$ transmit antennas and $M$ receive ones. The specification of the antenna is extracted from [23] which describes the design of a high-gain antenna for the D-band with transmit-arrays technology.

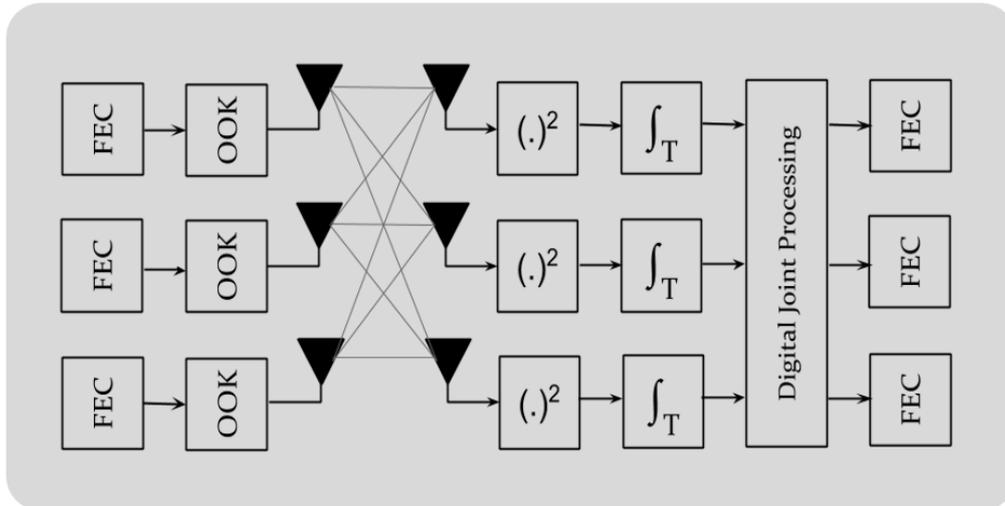

Fig. 9: System model for the D2D scenario.

To limit the complexity and the power consumption of the transceiver, we propose to use on-off keying (OOK) combined with a Bose, Ray-Chaudhuri and Hocquenghem (BCH) code. The considered BCH code has a packet size of 63 bits, a coding rate ranging from 0.4 to 1 and its decoder is based on hard decisions. The key features of this code are a low-complexity implementation and a low-power consumption. With regard to the short packet length, the resulting decoder also presents a low latency. These features appear to be highly relevant for the D2D applications. As depicted in Fig. 9, channel coding is implemented independently by antenna stream. This architecture is particularly suited to high-rate applications as it maintains a high degree of parallel processing.

By way of illustration, an 8 × 8 MIMO system can communicate over 5 m in the D-band with a data rate of 3.25 Gbps/ch. for a 1 GHz band, a BER below $10^{-6}$ and a transmit power by antenna $P_{ATx} \leq$ −43 dBm [22]. Therefore, we conclude that MIMO systems with ED receivers offer a valuable solution to achieve high-rate systems in sub-THz bands with extremely low-complexity and low-power RF architectures.

In order to achieve a higher data rate, a coherent system based on FSIM scheme can be used. For instance, the LDPC coded FSIM has been explored in the D2D scenario to highlight its advantages. The LDPC channel coding is considered a codeword length of 16200 bits, and it is applied on each antenna stream of Fig. 8 independently. For example, SISO and 8x8 MIMO FSIM using only 2 filters and QPSK with linear receiver require a transmit power $P_{A_{Tx}} = -48.3$ dBm/ch. and $P_{A_{Tx}} = -36$ dBm/ch for 5m distance and a maximum data rate of 2.67 Gbps/ch. and 21.33 Gbps/ch., respectively. The SISO-FSIM reduces the number of transceiver RF chains, and thus the cost. Note that higher data rates in SISO and MIMO can be achieved easily by increasing the number of filter shapes.

Therefore, all the analysis promotes the SISO/MIMO FSIM scheme as a very competitive and promising candidate for low-power wireless ultra-high data rates system in sub-THz bands. The reconfigurable FSIM based access point can serve the end-users of different constraint levels. For instance, transmitting FSIM with APM symbol is very convenient for low-power ultra-high data rate coherent systems, while switching to its special case OOK is recommended when communicating with an extremely low complexity/cost non-coherent receiver suffering from strong PN. Interested readers can refer to [8] [4], BRAVE deliverable [18], and references therein for more details.

# VIII. CONCLUSION

Although lots of research efforts have been undertaken with state-of-the-art devices and channel propagation in the last years, sub-THz techniques are still not mature enough compared to microwave or photonics technologies. The substantial studies and technical advances provided in this paper allow taking a step forward on developing realistic sub-THz communication systems.

To sum up the requirements, constraints, and results of the previously investigated scenarios, Table 3 reports the key performance indicators. The maximum achievable throughput is given for different PN conditions using 1 GHz channel bandwidth. The transmit power is given per channel and per antenna while taking into account the noise figure and implementation/interconnection losses.

Table 3: Key performance indicator for different sub-THz scenarios.

| Scenario | System description | Tx power per antenna | Antenna gains (Tx,Rx) | Losses + Noise Figure | Through. / 2GHz-channel | Range |
|---|---|---|---|---|---|---|
| Backhaul Beamforming | LDPC coded SISO Coherent P-QAM | 20 dBm | (32, 32) dBi | (3 + 10) dB | No PN: 5.7 Gbps<br>Strong PN: 4.4 Gbps | ~300 m |
| Short Range (Kiosk, enhanced Hotspot, etc.) | **Uncoded** 10×10 MIMO GSM-QPSK with 3 active TAs and joint ML detection | -6 to 14 dBm for uncoded BER<$10^{-4}$ | (10, 10) dBi | 12 dB | Medium PN: Up to 12 Gbps | 0.5 m to 5 m |
| | **Uncoded** 4x10 MIMO 2-FSIM-QPSK with linear receiver | -9.5 to 10.5 dBm for uncoded BER<$10^{-4}$ | (10, 10) dBi | 12 dB | Medium PN: Up to 12 Gbps | 0.5 m to 5 m |
| D2D | LDPC coded 8x8 MIMO 2-FSIM-QPSK with linear receiver | -56 to -36 dBm | (32, 32) dBi | 10 dB | Medium PN: Up to 21.33 Gbps | 0.5 m to 5 m |
| | LDPC coded SISO 2-FSIM-QPSK with linear receiver | -68 to -48.3 dBm | (32, 32) dBi | 10 dB | Medium PN: Up to 2.67 Gbps | 0.5 m to 5 m |
| | BCH coded 8X8 MIMO non-coherent OOK with energy detector | < -43 dBm | (32, 32) dBi | 10 dB | Strong PN: 3.25 Gbps | 5 m |

We demonstrate that optimizing the modulation and demodulation schemes for PN channels, i.e., with Polar-QAM, results in significant performance gains for future sub-THz systems. In addition, the combination of MIMO with spectral-efficient IM scheme and power-efficient SC modulations allows reaching an ultra-high throughput with low power consumption. This combination considers that the long-term evolution roadmap of the next-generation telecommunications systems (Beyond-5G) will point to an energy-efficiency dominated era. This approach shows many advantages compared to its predecessor in sub-GHz high data-rate systems (e.g., IEEE 801.11ax). Following this approach, SISO FSIM and different MIMO candidate systems (GSM, SMX, DP-GSM, DP-SMX, SMX-FSIM) are investigated, where FSIM based system with linear receiver is the most promising developed solution for low-power wireless Tbps. Last, we have highlighted that SISO-FSIM and MIMO-FSIM are also advantageous for D2D applications of low-complexity/cost. However, using MIMO systems with ED receivers is a valuable solution for an extremely low-cost/complexity receiver. It allows achieving a high and sufficient data-rate communications with low complexity/cost and low-power transceivers in sub-THz frequencies.

Accurate and smart radio-planning functions will be necessary for deployment of backhaul or local access networks due to the strong environmental constraints that are met at those frequencies. As shown in the backhaul and FWA scenario, same technology can be used to assess the feasibility and deployment expenditure at the research stage; this allows for early refinement of the target use cases and requirements.

These results sound here of significant advances in a path still full of challenges. Several technical challenges on power consumption, low complexity RF architectures, and compact antenna designs need to be overcome to realize the development and deployment of future wireless sub-THz communication devices.

White paper - Wireless connectivity in the sub-THz spectrum: A path to 6G[14] T. Xing and T. S. Rappaport, "Propagation Measurement System and Approach at 140 GHz– Moving to 6G and Above 100 GHz," in *2018 IEEE Global Communications Conference (GLOBECOM)*, Abu Dhabi, UAE, Dec. 2018.

[15] G. Gougeon, Y. Corre and M. Z. Aslam, "Ray-based Deterministic Channel Modelling for sub-THz Band," in *2019 IEEE International Symposium on Personal, Indoor and Mobile Radio Communications (PIMRC)*, Istanbul, Turkey, Sep. 2019.

[16] S. Bicaïs and J.-B. Doré, "Phase Noise Model Selection for Sub-THz Communications," in *2019 IEEE Global Communication Conference (GLOBECOM)*, Waikoloa, USA, Dec. 2019.

[17] G. Gougeon et al., "Assessment of sub-THz Mesh Backhaul Capabilities from Realistic Modelling at the PHY Layer," in *14th European Conference on Antennas and Propagation (EuCAP 2020)*, Copenhaguen, Denmark, Mar. 2020.

[18] BRAVE, "Performance assessment," BRAVE deliverable D3.1, Nov. 2021.

[19] J. Wang, S. Jia and J. Song, "Generalised spatial modulation system with multiple active transmit antennas and low complexity detection scheme," *IEEE Transactions on Wireless Communications,* vol. 11, no. 4, p. 1605–1615, Apr. 2012.

[20] M. Saad, F. C. Lteif, A. G. A. C., H. Hijazi, J. Palicot and F. Bader, "Generalized Spatial Modulation in Highly Correlated Channels," in *30th IEEE International Symposium on Personal, Indoor and Mobile Radio Communications (PIMRC 2019)*, Istanbul, Turkey, Sep. 2019.

[21] N. Bouhlel, M. Saad and F. Bader, "Sub-Terahertz Wireless System using Dual-Polarized Generalized Spatial Modulation with RF Impairments," *IEEE Journal on Selected Areas in Communications,* vol. 39, no. 6, pp. 1636-1650, Jun. 2021.

[22] S. Bicaïs, J.-B. Doré and V. Savin, "Design of MIMO Systems using Energy Detectors for Sub-TeraHertz Applications," in *IEEE International Symposium on Personal, Indoor and Mobile Radio Communications (PIMRC 2020)*, Sep. 2020.

[23] F. F. Manzillo, A. Clemente and J. L. Gonzalez-Jimenez, "High-gain D-band Transmitarrays in Standard PCB Technology for Beyond-5G Communications," *IEEE Transactions on Antennas and Propagation,* vol. 68, no. 1, pp. 587-592, Jan. 2020.
25

# ABOUTS

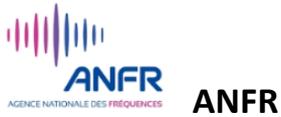 **ANFR**

**ANFR** (Agence nationale des fréquences) is the French governmental institution that is responsible of the RF spectrum management: spectrum planning, frequency allocation, coordination of frequency assignments, control and international negociations.

**ANFR** is providing assistance in understanding the spectrum regulation constraints and opportunities.

**Contact person:** Emmanuel Faussurier, mailto:emmanuel.faussurier@anfr.fr.

Website: https://www.anfr.fr/accueil/.

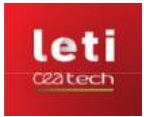 **CEA-Leti**

**Leti**, a technology research institute, pioneers micro and nanotechnologies, tailoring differentiating applicative solutions that ensure competitiveness in a wide range of markets. The institute tackles critical challenges such as healthcare, energy, transport and ICTs. Its multidisciplinary teams deliver solid expertise for applications ranging from sensors to data processing and computing solutions, leveraging world-class pre-industrialization facilities. Leti builds long-term relationships with its industrial partners – global companies, SMEs and startups – and actively supports the launch of technology startups.

**Leti** is part of CEA tech, the technology research branch of the French Alternative Energies and Atomic Energy Commission (CEA). It is a key player in research, development and innovation in defense & security, nuclear energy, technological research for industry and fundamental physical and life sciences. In 2015, Thomson Reuters identified CEA as the most innovative research organization in the world.

**CEA-Leti** has strong know-how on future mmW systems that is used to identify new challenges for upcoming research activities. It is contributing to the BRAVE project to impact the Beyond 5G research and perspectives by the evaluation of several candidate scenarios and technologies.

**CEA-Leti** is contributing the RF impairment modelling, and is leading the third workpackage on integration and simulation.

**Contact person:** Jean-Baptiste Doré, jean-baptiste.dore@cea.fr.

**Website:** https://www.cea.fr/cea-tech/leti/.

**CentraleSupélec**

**CentraleSupélec** is a public institution under ministerial charter, devoted to the sciences and engineering. This charter is shared between the Ministry of Higher Education, Research and Innovation, and the Ministry of Economy, Industry and Digital Technologies. CentraleSupélec was officially established on January 1st, 2015, bringing together two leading engineering schools in France: Ecole Centrale Paris and Supélec.

**CentraleSupélec** research center is composed of 17 laboratories and research teams which run in co-operation with major national research centers; 1 research federation in mathematics; 1 research institute with EDF; 4 international laboratories with China, Canada, The United States and Singapore; 1080 staff including 300 faculty researchers, 65 full-time researchers, 600 PhD candidates, 70 postdoctoral students, and 145 administrative and technical staff; and 17 research chairs.

The **SCEE** (Signal, Communication & Embedded Electronics) research group is a research team of the IETR (Institute of Electronics and Telecommunications of Rennes - UMR CNRS 6164) that contributes to the BRAVE project belongs to CentraleSupélec campus of Rennes.

**Contact persons:** Carlos Faouzi Bader, mailto:carlos.bader@centralesupelec.fr, Majed Saad mailto:majed.saad@ieee.org

**Website:** http://www.rennes.centralesupelec.fr/en/home.

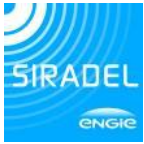 **SIRADEL**

**SIRADEL** is an innovative high-tech company that helps cities to be better connected and sustainable. It provides solutions for wireless networks design and urban transformation planning based on a unique combination of technologies and know-how in the field of 3D city modeling, simulation and 3D visualization. Founded in 1994, based in France (HQ), Canada and China, a subsidiary of ENGIE since 2016, with 130 employees, SIRADEL operates with over 250 key account clients (telecoms operators and equipment manufacturers, municipalities, energy suppliers, transport companies, and national and international authorities) in more than 60 countries.

**SIRADEL** solutions optimize cost and performance of the connectivity networks (cellular 3G/4G/5G NR, IoT, WiFi, satellites, small cells, millimetre-wave access and backhaul, fiber, …), energy networks (smart grid, heating and cooling systems), smart lighting networks (LED, LIFI), geolocation-based networks (GNSS, indoors) or assets for security (cameras, sensors) or smart mobility (autonomous vehicles, drones). It contributes to accelerate the convergence between wireless, fixed networks and all urban infrastructures and brings solutions to connect people, objects and assets. SIRADEL participates in several public and industrial research programs in future smart-cities design, aimed at the simulation and optimization of next generation urban infrastructures.

**SIRADEL** has upgraded its propagation channel models and high-frequency network simulators to sub-THz bands, in order to contribute to the evaluation and demonstration of beyond-5G communication scenarios.

**Contact person**: Yoann Corre, ycorre@siradel.com, coordinator of the BRAVE project.

**Website:** https://www.siradel.com.

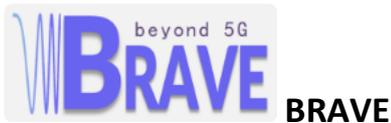 **BRAVE**

**BRAVE** is a collaborative research project started in January 2018, that aims at creating new physical-layer (PHY) techniques devoted to beyond-5G wireless communications. The partners (Siradel, CentraleSupélec, CEA-Leti and ANFR) are designing new high-data-rate and energy-efficient waveforms that operate in frequencies above 90 GHz. Application to scenarios such as kiosks, backhauling, hotspots are assessed to evaluate the benefit of the proposed technology.

Funded by the French Research Agency ANR, the **BRAVE** project will go to its end in September 2021.

The consortium is combining skills from an industry (Siradel), an academic laboratory (Centrale-Supélec), a research institute (CEA-Leti) and a regulator (ANFR), which are all familiar with collaborative research and recognized in the field of wireless innovation. The partners bring complementary skills to efficiently tackle the different challenges of the project: regulation, signal processing, realistic modelling of the PHY-layer, and software-based evaluation.

The **BRAVE** partners are exploring a new radio technology that is intended to operate at high frequencies (above 5G spectrum), along with huge bandwidth, and enhanced spectral and energy efficiencies. Ultimate goal is the definition of a solution that would reach 1 Tbps.

**Website:** http://www.brave-beyond5g.com/.

**All publications and deliverables are available at**:
http://www.brave-beyond5g.com/index.php/publications/.